 \documentclass[12pt]{article}
 \usepackage{amssymb}

 \newcommand{\R}{ {\mathbb R} }

\begin{document}

 \vspace{10pt}

 \begin{center}
 \large\bf

 Spherically-symmetric solutions with a chain of $n$  internal
 Ricci-flat spaces

 \vspace{15pt}

 \normalsize\bf
         V. D. Ivashchuk

 \vspace{10pt}

 \it  \ \ \ Center for Gravitation and Fundamental
 Metrology,  VNIIMS, 46 Ozyornaya Str., Moscow 119361, Russia  \\

 Institute of Gravitation and Cosmology,
 Peoples' Friendship University of Russia,
 6 Miklukho-Maklaya Str.,  Moscow 117198, Russia \\

 \end{center}
 \vspace{10pt}

  \begin{abstract}
 The Schwarzschild solution is generalized
 for the case of $n$ internal Ricci-flat spaces. It is shown
 that in the four-dimensional section of the metric a horizon
 exists only when the internal space scale factors are constant.
 The scalar-vacuum generalization of the solution is also
 presented.
 [This paper is the English translation of the part of  Chapter. 2.4
  of the author's PhD dissertation (Moscow, 1989).]

 \end{abstract}

\vspace{5pt}

\setcounter{equation}{0}

 Here we derive an exact  solution to vacuum Einstein equations
 $R_{MN} = 0$ in a  spherically-symmetrical case when all internal spaces
 $M_1, \dots, M_n$ are  Ricci-flat \cite{BrIvMel-89,Ivashchuk-PhD}.

 So, the problem is to find a solution for the metric of the form

 \begin{eqnarray}
 g =&-&e^{2\gamma (u)}dt\otimes dt + e^{2\alpha (u)}du \otimes du +
 \nonumber \\
 &+& e^{2\beta_0(u)}d\Omega^2 +
 \sum_{i=1}^{n}e^{2\beta_i(u)}g_{(i)} \label{1}
 \end{eqnarray}
  on the manifold
 \begin{equation} \label{2}
 M = \R \times \R_{*} \times S^2 \times M_1 \times \cdots \times M_n,
 \end{equation}
obeying the vacuum Einstein eqs., where $M_i$ is a Ricci-flat
manifold of dimension $N_i$ with the metric $g_{(i)}$,
 $i=1,...,n$,  $d\Omega^2$ is the canonical metric on $S^2$,
 $ \R_{*} \subset \R $ and  $u$ is a
radial-type variable connected with $r$ by the relation
$r=e^{\beta_0(u)}$. Denote $\gamma =\beta_{-1}$, $N_{-1}=1$,
$N_0=2$. Let $\alpha =\alpha_0\equiv \sum_{\nu
=-1}^{n}\beta_{\nu}N_{\nu}$ ($u$ is a harmonic radial variable).

 Then Einstein eqs. $R_{MN}=0$  read $(A'\equiv \frac{d}{du}A)$

 \begin{eqnarray}
 &&\sum_{\nu
 =-1}^{n}[-\beta_{\nu}^{''}+\alpha_0'\beta_{\nu}'-(\beta_{\nu}')^2]N_{\nu}=0,
 \nonumber \\
 && \beta_{i}^{''} = 0,\ \  i=-1,1,\cdots ,n,  \label{3} \\
 && \beta_{0}^{''} = e^{2\alpha_0-2\beta_0}. \nonumber
 \end{eqnarray}

 Solving the set of equations (\ref{3})
 (here it is convenient to use the  variable $x=\beta_0-\alpha_0$) we get

 \begin{eqnarray}
 \beta_i&=& A_i\overline{u}+D_i, \ \ i=-1,1,\cdots ,n, \nonumber  \\
 \beta_0 &=& - \ln f-\sum_{\nu \neq 0}(A_{\nu}\overline{u}+D_{\nu})N_{\nu},
  \label{4} \\
  \alpha_0 &=& -2 \ln f-\sum_{\nu \neq
  0}(A_{\nu}\overline{u}+D_{\nu})N_{\nu}, \nonumber
  \end{eqnarray}
where
 \begin{equation}
  f=f(\overline{u},B) = \left \{
 \begin{array}{lc}
 \frac{sh(\sqrt{B}\overline{u})}{\sqrt{B}}, & B>0,    \\
 \overline{u}, & B=0.
 \end{array}
  \right.
 \end{equation}

In (\ref{4}) $\overline{u}=\varepsilon (u+u_0), \varepsilon =\pm
 1; u_0, A_i, D_i$ are arbitrary constants, $i=-1,1,\cdots ,n$. B
is defined by the relation
 \begin{equation}
 2B = \left(\sum_{\nu \neq 0}A_{\nu}N_{\nu}\right)^2 + \sum_{\nu
 \neq 0}N_{\nu}A_{\nu}^2 \label{6}
 \end{equation}
 $(\displaystyle\sum_{\nu \neq 0}$ means the summation over $\nu$ : $\nu
  =-1,1,\cdots ,n$). If we re-denote the constants

 \begin{eqnarray}
 && c_i=e^{2D_i}, \ \ a_i\sqrt{B}=-A_i, \ \ i=1,\cdots ,n; \nonumber \\
 && c=e^{D_{-1}}, \ \ a\sqrt{B}=-A_{-1}, \\
 && L=2\sqrt{B} \exp(-\sum_{\nu \neq 0}D_{\nu}N_{\nu})
 \nonumber
 \end{eqnarray}

  and introduce a new variable $R$:
 \begin{equation}
  R = e^{-\sum_{\nu \neq 0}D_{\nu}N_{\nu}}\times \left \{
 \begin{array}{lc}
 \frac{2\sqrt{B}}{1-e^{-2\overline{u}\sqrt{B}}}, & B>0,
 \\
 1/\overline{u}, & B=0,
 \end{array}
 \right.
 \end{equation}
then the solution (\ref{1}),  (\ref{4}) reads
\cite{BrIvMel-89,Ivashchuk-PhD}

\begin{eqnarray}
 &&g=-c^2dt\otimes dt\left(1-\frac{L}{R}\right)^a + dR\otimes
 dR\left(1-\frac{L}{R}
 \right)^{-a-\sum_{i=1}^{n}a_iN_i} + \nonumber \\
 &&+d\Omega^2R^2\left(1-\frac{L}{R}\right)^{1-a-\sum_{i=1}^{n}a_iN_i}
 + \sum_{i=1}^{n} c_ig_{(i)}\left(1-\frac{L}{R}\right)^{a_i},
 \label{9}
 \end{eqnarray}
$R>L$, where constants $L\geq 0,c,c_1,\cdots ,c_n>0$ are arbitrary
and $a,a_1,\cdots ,a_n$ obey the relation following from (\ref{6})
 \begin{equation}
  \left(a+\sum_{i=1}^{n}a_iN_i\right)^2 +
  a^2+\sum_{i=1}^{n}a_i^2N_i=2.  \label{10}
 \end{equation}

 Solution (\ref{9}), (\ref{10}) for special case $n=1$ was considered earlier in
 \cite{Yoshimura, Myers}.
 [For one-dimensional internal spaces see also \cite{Kramer,Lyogkiy}
  ($n=1$) and \cite{Vladimirov}($n=2,3$).]

When $L=0$ the solution (\ref{9}) is trivial: in this case
4-dimensional part of the metric (\ref{9})  is flat and scale
factors for $g_{(i)}$ are constant. For $L>0$ and

\begin{equation}
 a-1=a_1=\cdots =a_n=0 \label{11}
\end{equation}
the solution (\ref{9}) is the sum of the Schwarzschild solution
(with the gravitational radius $L$) and the tensor field
$\sum_{i=1}^{n}c_ig_{(i)}$. Let $L>0$, then  $a>0$ corresponds to
an attraction and $a<0$ describes a repulsion.

Now let us study the problem of a horizon for the solution
(\ref{9}). Consider $g_4$ which is  the 4-dimensional section of
the metric (\ref{9}). For the metric $g_4$ in the non-trivial case
$L>0$ the horizon  at $R=L$ takes place only when (\ref{11})
holds. Indeed, for a  radial light geodesic obeying $ds_4^2=0$ we
have
\begin{equation}
 c(t-t_0) =
 \displaystyle\int_{R}^{R_0}dx
 \left(1-\frac{L}{R}\right)^{-a-\frac{1}{2}\sum_{i=1}^{n}a_iN_i}.
  \label{12}
\end{equation}

Relation (\ref{10}) is equivalent to the following identity
 \begin{equation}
 \left(a+\frac{1}{2}\sum_{i=1}^{n}a_iN_i\right)^2 =
 1-\frac{1}{2}\sum_{i=1}^{n}a_i^2N_i
 -\frac{1}{4}\left(\sum_{i=1}^{n}a_iN_i\right)^2. \label{13}
 \end{equation}

If not all $a_i=0$ ($i=1,\cdots ,n)$, then due to (\ref{13})
 \begin{equation}
 \mid a+\frac{1}{2}\sum_{i=1}^{n}a_iN_i\mid <1,
 \end{equation}
and so the integral (\ref{12}) is convergent for $R=L$, i.e. a
radial light ray reaches the surface $R=L$ at a finite time. If
 $a_1=\cdots =a_n=0$ then due to (\ref{10}) $a=\pm 1$. When $a=1$,
 $a_1=\cdots =a_n =0$ the metric $g_4$ coincides with the Schwarzschild
solution having a horizon at $R=L$. If $a=-1$, $a_1=\cdots =a_n=0$
then the integral (\ref{12}) is  finite for $R = L$ and hence the
horizon is absent. Thus, for the metric $g_4$ (which is  the
4-dimensional section of the metric (\ref{9})) the surface $R=L$
is a horizon only in the trivial case (\ref{11}) when scale
factors of internal spaces are constant and 4-section of the total
metric coincides with the Schwarzschild solution.

 Solution (\ref{9})  could be easily generalized when a
 minimally coupled scalar field   is taken into account.
 In this case  the action of the model
 \begin{equation}
 S= \frac{1}{2}\int d^D z \mid
 g\mid^{1/2}\left(\frac{R[g]}{\kappa^2}-g^{M N}
 \partial_{M}\varphi \partial_{N}\varphi \right) \label{15}
 \end{equation}
leads to equations of motion
 \begin{eqnarray}
 R_{MN} &=& \kappa^2\partial_M\varphi \partial_N\varphi,  \label{16}\\
 \Delta \varphi &=& 0, \label{17}
 \end{eqnarray}
where $\Delta$ is the Laplace operator for the metric $g$. For the
metric (\ref{9}) and the scalar field $\varphi = \varphi(u)$
(where $u$ is a harmonic radial variable) eq. (\ref{17}) reads:
 $\varphi'' = 0$, or, equivalently,
 \begin{equation}
 \varphi = Q u+ \bar{\varphi}_0, \label{18}
 \end{equation}
 where $Q$ and $\bar{\varphi}_0$ are constants. The
 substitution of the metric (\ref{1}) and the scalar field from
(\ref{18}) into eqs. (\ref{16}) leads us to a set of equations
 which differs from (\ref{3}) by the presence of  the term
 $\kappa^2 Q^2$ in the right hand side of the first equation
 of the set (\ref{3}).  Solving this modified set of equations
 (along a line as it was  done for the set (\ref{3})) we get
\begin{equation}
  \varphi = \frac{1}{2}q \ \ln \left(1-\frac{L}{R}\right)+ \varphi_0, \label{19}
\end{equation}
where $q$ and $\varphi_0$  are constants ($q$ is scalar charge),
and  the metric $g$ is given by the same formula (\ref{9}) but
instead of (\ref{10})  the constants $a,a_1,\dots, a_n$ obey the
relation \cite{Ivashchuk-PhD}

\begin{equation}
 \left(a+\sum_{i=1}^{n}a_iN_i\right)^2+a^2+\sum_{i=1}^{n}a_i^2N_i+\kappa^2q^2=2
 \label{20}.
\end{equation}

The scalar-vacuum solution for $n=1$ was considered earlier in
\cite{BrIv-89}.  It is easy to prove  using (\ref{20}) that a
horizon for $R=L > 0$ takes place only when

\begin{equation}
 q=a-1 = a_1=\cdots =a_n=0.
\end{equation}


 \end{document}